\documentclass[aps,prb,twocolumn,showpacs,amsmath,preprintnumbers,amssymb,superscriptaddress,10pt]{revtex4}

\usepackage{graphicx}
\usepackage{dcolumn}
\usepackage{bm}
\usepackage[latin1]{inputenc}
\usepackage{color}
\begin{document}

\newcommand{\ittext}[1]{\mbox{\rm\scriptsize #1}}
\title{Energy Shift and Wavefunction Overlap of Metal-Organic Interface-States}

\author{M.~Marks}
\affiliation{Fachbereich Physik und Zentrum f{\"u}r Materialwissenschaften,
Philipps-Universit{\"a}t, 35032 Marburg, Germany}

\author{N.~L.~Zaitsev}
\affiliation{Donostia International Physics Center (DIPC), 20018 San
Sebasti{\'a}n, Spain}
 \affiliation{Tomsk State University, 634050, Tomsk, Russia.}

\author{B.~Schmidt}
\affiliation{Fachbereich Physik und Zentrum f{\"u}r Materialwissenschaften,
Philipps-Universit{\"a}t, 35032 Marburg, Germany}

\author{C.~H.~Schwalb}
\affiliation{Fachbereich Physik und Zentrum f{\"u}r Materialwissenschaften,
Philipps-Universit{\"a}t, 35032 Marburg, Germany}

\author{A.~Sch{\"o}ll}
\affiliation{Universit{\"a}t W{\"u}rzburg, Experimentelle Physik~VII
and Röntgen Center (RCCM), 97074 W{\"u}rzburg, Germany}

\author{I.~A.~Nechaev}
\affiliation{Tomsk State University, 634050, Tomsk, Russia.}
\affiliation{Nekrasov Kostroma State University, 156961 Kostroma, Russia.}

\author{P.~M.~Echenique}
\affiliation{Donostia International Physics Center (DIPC), 20018 San
Sebasti{\'a}n, Spain}
\affiliation{Departamento de F{\'\i}sica de Materiales, UPV/EHU
and CFM - MPC, Centro Mixto CSIC - UPV/EHU,
20080 San Sebasti\'an, Spain}

\author{E.~V.~Chulkov}
\affiliation{Donostia International Physics Center (DIPC), 20018 San
Sebasti{\'a}n, Spain}
\affiliation{Departamento de F{\'\i}sica de Materiales, UPV/EHU
and CFM - MPC, Centro Mixto CSIC - UPV/EHU,
20080 San Sebasti\'an, Spain}

\author{U.~H{\"o}fer}
\affiliation{Fachbereich Physik und Zentrum f{\"u}r Materialwissenschaften,
Philipps-Universit{\"a}t, 35032 Marburg, Germany}
\affiliation{Donostia International Physics Center (DIPC), 20018 San
Sebasti{\'a}n, Spain}

\date{20.06.2011}

\begin{abstract}

The properties of Shockley-type interface states between
$\pi$-conjugated organic molecular layers and metal surfaces are
investigated by time-resolved two-photon photoemission experiments
and density functional theory. For perylene- and
naphthalene-tetracarboxylic acid dianhydride (PTCDA and NTCDA)
adsorbed on Ag(111), a common mechanism of formation of the
interface state from the partly occupied surface state of the bare
Ag(111) is revealed. The energy position is found to be strongly
dependent on the distance of the molecular carbon rings from the
metal and their surface density. Bending of the carboxyl groups
enhances the molecular overlap of the interface state.


\end{abstract}

\maketitle


The energetic position and wavefunction overlap of electronic states
at the interface between layers of organic molecules and metals is
of fundamental interest for the design of organic semiconductor
devices and for future applications of molecular electronics.
Previous studies, both experimentally and theoretically,
concentrated on those electronic states that either result directly
from the chemical bonding at the interface or from the shift and
broadening of localized molecular orbitals upon interaction with the
metal
substrate.\cite{Hwang09,Koch08jp,Rohlfing07prb,Zou06ss,Zhu04ssr,Tegeder07apa,Yamane07prb,Gonzal08prl}
However, also states intrinsic to metal surfaces are affected by
adsorption and can become an important factor for the electronic
coupling between metal and organic molecules.
 This has recently become apparent for
3,4,9,10-perylene-tetracarboxylic-acid-dianhydride (PTCDA) on
Ag(111), a structurally very well characterized model system for the
interface between $\pi$-conjugated organic molecular layers and
metals.\cite{Zou06ss}
 A dispersing, free-electron-like electronic state located 0.6 eV
above the Fermi level was observed by tunnelling spectroscopy
\cite{Temirov06nat} and by two-photon photoemission
(2PPE).\cite{Yang08jpcc,Schwalb08prl}
 It originates from the Shockley state of
the Ag(111) surface which is upshifted from below the metallic Fermi
level by as much as 0.7 eV. The initially partially occupied state
becomes unoccupied and approaches the conduction band of the organic
semiconductor.\cite{Schwalb08prl,Zaitsev10,Dyer10njp}

In this Letter, we address the question which properties of the
molecule-surface interaction determine the energetic position of
such Shockley-type interface states (IS) and which factors
facilitate a large overlap of the state with the molecular layer in
order to tailor the degree of electronic coupling.
 For rare-gas adsorbates on noble metals it is well established that
their adsorption also leads to an upshift of the Shockley state,
albeit by much smaller values, and that the shift increases
systematically from the weakly bound Ar to the more strongly
interacting Xe.\cite{Andreev04prb,Forster04jpcb}
 The comparison of our 2PPE experiments for
1,4,5,8-naphthalene-tetracarboxylic acid dianhydride (NTCDA) on
Ag(111) with PTCDA/Ag(111) shows an analogous trend. The structurally
similar, but smaller NTCDA molecules cause a smaller IS upshift than
the larger PTCDA molecules. The time-resolved 2PPE data indicate a
similar overlap of the wavefunction with the metal in both cases.
 Density functional calculations explain the smaller shift by a
larger distance of the NTCDA carbon plane from the surface.
 Surprisingly, in the metastable disordered PTCDA phase, when the molecules
interact more strongly with the substrate,\cite{Kilian08prl} we
observe a smaller and not a larger IS upshift.
 Lifetime measurements indicate a reduced overlap of the wavefunction
with the metal.
 Theory relates this finding to the bending of the carboxyl groups
which enables the IS wavefunction to extend more into the molecule.

\begin{figure*}[tb]
 \includegraphics[width = 0.9\textwidth]{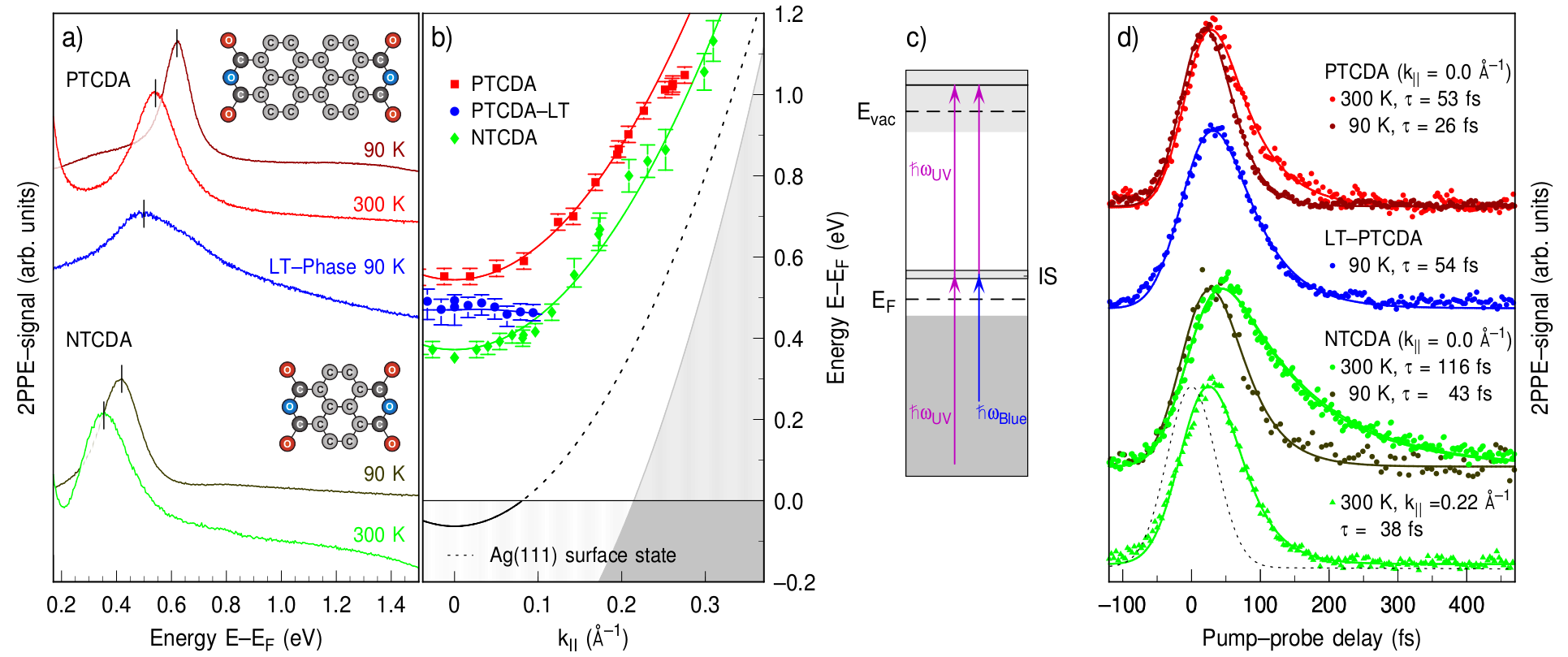}
\caption{
 Energy-, angle- and time-resolved 2PPE data of monolayer films of
NTCDA, PTCDA on Ag(111).
 (a) Normal emission (UV+UV) 2PPE spectra of commensurate PTCDA and NTCDA
monolayers recorded at 300~K and 90~K as indicated, and of the
disordered LT-phase of PTCDA at 90~K.
 The insets sketch the molecular structure for PTCDA (top) and NTCDA
(bottom).
 (b) Angular dependence of the peak maxima of the interface state of
the ordered PTCDA (squares, red) and NTCDA (diamonds, green) phases
measured at 300~K and the PTCDA low-temperature phase (dots, blue)
measured at 90~K together with projected Ag(111) bulk bands (gray
shaded area) and the Shockley state of the clean surface (black).
 (c) Excitation schemes employed to 2PPE spectroscopy (left) and
time-resolved 2PPE (right).
 (d) Time-resolved 2PPE data recorded at the peak maxima of the
interface states of the different PTCDA and NTCDA monolayers.
Lifetimes were extracted from best fits of a rate equation model
using Gaussian laser pulses that reproduce the measured
(Blue+UV)-cross-correlation on the clean surface (dashed line).
 \label{2ppe_figure}}
\end{figure*}


The experiments have been conducted under ultra-high vacuum
conditions using a 2PPE setup described previously.\cite{Sachs09jcp}
The photon energies of the blue and ultraviolet (UV) laser-pulses
were 3.10~eV and 4.70~eV with pulse lengths of 47~fs and 70~fs,
respectively. The overall energy resolution of the setup was 65~meV,
the angular resolution $1.2^{\circ}$.
 Thin organic films of temperature gradient sublimated PTCDA and
NTCDA were deposited onto sputter/annealed, clean Ag(111) surfaces
at a rate of 0.2--0.4~ML/min. Thermal desorption of multilayer films
resulted in a commensurate monolayer for PTCDA \cite{Zou06ss} and
the relaxed monolayer for NTCDA.\cite{Stahl98ss} Sample cooling to
90~K was performed rapid enough ($\gtrsim$ 15~K/min) to suppress the
inverse melting phenomenon for NTCDA.\cite{Scholl10sci} The PTCDA-LT
phase was prepared by adsorption at a sample temperature of 90~K.

The DFT calculations have been performed using the pseudopotential
method as implemented in OPENMX code within the local density
approximation for the exchange-correlation
potential.\cite{Ozaki05prb,Zaitsev10}
 A slab containing a 9-layer silver film was employed. The relaxed
NTCDA monolayer was attached on one side of the film, the vacuum
region was chosen to correspond to 6 silver interlayer spacings.
 The resulting periodic supercell has a rectangular unit cell of
$11.56\times15.02$~\AA$^2$ and contains 216 silver atoms and 48
atoms belonging to two NTCDA molecules.\cite{Stahl98ss} The surface
Brillouin zone (SBZ) was sampled with a $3\times3$
$\mathbf{k}$-point mesh.


Figs.~1(a,b) display the results of angle-resolved 2PPE spectroscopy
performed with a single-color (UV+UV)-excitation scheme (cf.
Fig.~1(c)).
 In agreement with previous results, the well-ordered herringbone
phase of PTCDA/Ag(111) exhibits a strongly dispersing state $\sim
0.6- 1.0$ eV above the Fermi level ($E_{\rm F}$).\cite{Schwalb08prl}
For one monolayer the band minimum at $k_\|=0$ is \mbox{$E - E_{\rm
F} = 0.57\pm0.02$~eV} at room temperature. Its energy increases
slightly when additional layers are adsorbed or when the sample is
cooled to liquid nitrogen temperature.\cite{Sachs09jcp}
 Similar to PTCDA, the smaller NTCDA molecules can be adsorbed in a
commensurate monolayer on Ag(111).\cite{Stahl98ss} We observe an
unoccupied dispersing interface state also for this system (green
lines and data points in Figs.~1(a,b)). It is located 0.2~eV closer
to the Fermi level  \mbox{($E - E_{\rm F} = 0.38\pm0.02$~eV} at
$k_\|=0$) than the IS of PTCDA.
 The effective masses at the $\overline\Gamma$-point of both IS are
similar (PTCDA: $0.46\pm0.1\,m_e$,  NTCDA: $0.47\pm0.1\,m_e$) and
both values are close to that of the Shockley state of the clean
metal\cite{Forster04jpcb} ($0.42\,m_e$).


\begin{figure*}[bt]
  \includegraphics[width = 0.8\textwidth]{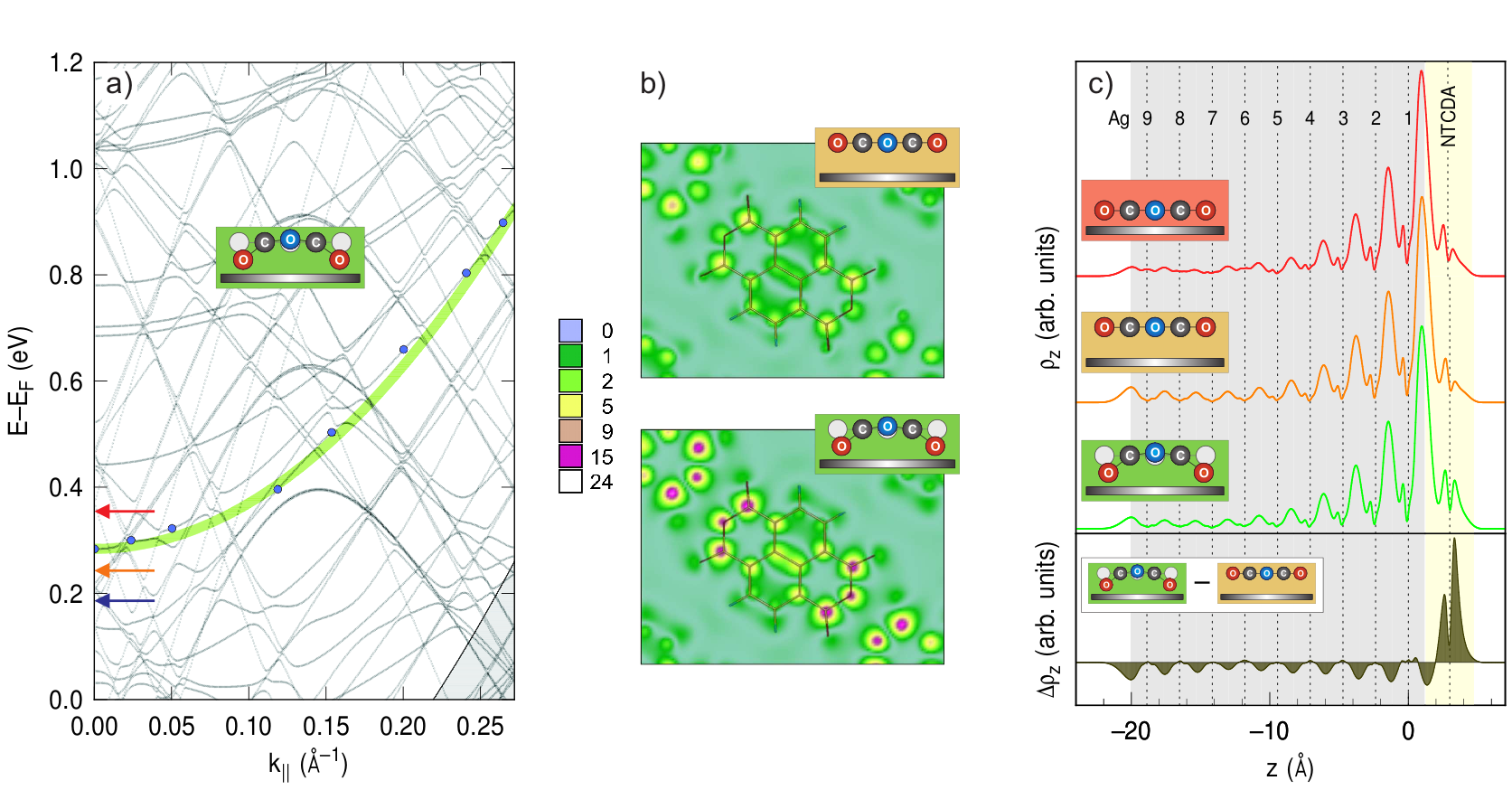}
\caption{
 (a) Band structure of the NTCDA monolayer on the nine-layer Ag(111)
film in the $\overline{\Gamma}\rightarrow\overline{X}_1$ direction
of the surface Brillouin zone for the bend molecular geometry.
 Filled circles indicate wave vectors $\mathbf{k}$, at which the spatial
distribution of the IS wave function was analyzed, the thick line
represents the parabolic approximation of the corresponding
dispersion, arrows indicate IS energies at $\overline{\Gamma}$ for
different geometries (from top to bottom: red: $d=2.85$~\AA-flat,
orange: $d=2.997$~\AA-flat, blue: less dense packing).
 (b) Spatial distribution of the IS probability density $\rho_{xy}$
at the $\overline{\Gamma}$-point in the vicinity of the molecular
plane for the flat (top) and the bent-down geometry (bottom).
 (c) IS probability density $\rho_z$ perpendicular to the surface,
$xy$-averaged over the unit cell (top: $\rho_z$ for the three
considered NTCDA geometries; bottom: difference $\Delta\rho_z$
between the bent and the flat geometry at the distance of 2.997~\AA.
 \label{dft_bandsstructure}}
\end{figure*}

The results of the time-resolved 2PPE experiments (Fig.~1(d))
underline the similarities between the IS of the commensurate PTCDA
and NTCDA layers.
 For these experiments we made use of a two-color (blue+UV)-excitation
scheme. The unoccupied IS is populated by an intense blue pump pulse
and the transient population is photoemitted by a weaker UV pulse
with variable time delay (Fig.~1(c)).\cite{fn1}
 The lifetime of $53\pm3$~fs determined for the IS of
the PTCDA herringbone phase at 300~K and $k_\|=0$ agrees well with
our previous results.\cite{Schwalb08prl} For NTCDA we measure a
considerably longer lifetime of $115\pm10$~fs.

Before comparing these two values we note that both IS are located
in the gap of the projected bulk bands of Ag(111). Electrons excited
into these states will decay inelastically by electron-hole-pair
creation at a rate determined by the wavefunction overlap with the
metal, the screening of the Coulomb interaction and the available
phase space in the metal.\cite{Echeni04ssr}
 Under the realistic assumption of a similar screened Coulomb
interaction at both interfaces (NTCDA/Ag and PTCDA/Ag) and taking
into account the different phase-space factors for the decay at the
different energies above the Fermi level, the measured lifetimes of
the IS allow us to compare their overlap with the metal.
 In fact, the ratio of the lifetimes
 \mbox{$\tau_{\rm NTCDA}/\tau_{\rm PTCDA}$} \mbox{[$116/53 = 2.19$]}
 is very close to the inverse phase space factor
 \mbox{$(E_{\rm PTCDA} - E_{\rm F})^2/(E_{\rm NTCDA}- E_{\rm F})^2$} \mbox{[$(0.57)^2/(0.38)^2 =
 2.25$]}
 of a free electron gas close to the Fermi level.\cite{Quinn58pr}
 This suggests a similar overlap of both states with the metal.
 We arrive at the same conclusion when we compare the lifetimes of the
interface states of NTCDA and PTCDA along their dispersing
bands.\cite{Berthold02prl}
 At $k_\|=0.22$~\AA$^{-1}$, e.g., the NTCDA-IS has a similar energy
as the PTCDA-IS at $k_\|=0.19$~\AA$^{-1}$ ($E - E_{\rm F} \simeq
0.8$~eV) and its lifetime (38~fs, Fig.~1(d)) is close to that of the
PTCDA-IS\cite{Sachs09jcp} (33~fs)).

Our results thus indicate that the commensurate NTCDA and PTCDA
layers have a qualitatively very similar influence on the Shockley
surface state of Ag(111), except for a larger upshift induced by the
more strongly interacting PTCDA.

 This picture changes when we turn to the results obtained for the
low temperature phase of PTCDA. Adsorption at sample temperatures
below 150~K leads to an incommensurate PTCDA monolayer in which the
molecules form dendritic islands. The reduced lateral interaction of
the molecules in this phase goes along with a stronger binding to
the surface.\cite{Kilian08prl}
 Surprisingly, this stronger binding is not accompanied by a larger
IS upshift as compared to the commensurate PTCDA phase. It is
observed at an energy of $0.50\pm0.03$~eV,\cite{fn2} i.e.\ 0.07~eV
{\em lower} than the IS of the more weakly bound commensurate phase
(Fig. 1(a)).
 The electron lifetime is $54\pm3$~fs (Fig. 1(d)).
 Since the IS lifetime of the commensurate PTCDA phase is strongly
temperature dependent,\cite{Sachs09jcp} we have to compare this
value, obtained at 90 K, with lifetime measurements of the
commensurate phases performed at the same temperature (Fig. 1(d)).
 The lifetime of $54\pm3$~fs is almost twice the 26~fs measured for
commensurate PTCDA at 90~K and even longer than the IS lifetime of
NTCDA (43~fs at 90~K) though the latter state is closer to the Fermi
level.
 Following the arguments above, the time-resolved experiments thus
indicate that the wave function of the IS of the low-temperature
PTCDA phase has less overlap with the metal than that of the ordered
PTCDA and NTCDA layers.


 In order to obtain a microscopic understanding of the origin of the
observed energy shifts of the IS and its wavefunction overlap, we
performed DFT calculations for three different geometries of the
NTCDA monolayer.
 First, the molecules stay in their flat gas phase geometry,
 while they are artificially kept at the experimentally determined
 adsorption distance of the carbon plane ($d=2.997$~\AA).
 Second, the flat molecules are moved slightly
closer to the surface ($d=2.85$~\AA), i.e.\ to the adsorption
distance of the ordered PTCDA monolayer.\cite{Kilian08prl} Third,
the experimental adsorption geometry of NTCDA with bent carboxylic
oxygen atoms is simulated ($d=2.997$~\AA).\cite{Stadler07njp}
 Comparing calculations for one molecule (NTCDA) in different
geometries allows us to pinpoint the relevant factors that determine
properties of the interface state very clearly, although we can of
course not expect full quantitative agreement with our experimental
results of NTCDA and PTCDA molecules in different adsorption
geometries.

Fig.~2(a) shows the calculated band structure of NTCDA on Ag(111) in
the experimentally observed geometry.
 Due to the presence of the molecular monolayer, the period of the
unit cell has grown in comparison with the clean Ag(111) surface,
and the corresponding surface Brillouin zone has reduced its size.
As a result, bands in the SBZ of the clean-surface unit cell get
``folded'' into the reduced SBZ. This leads to a surface band
structure that no longer exhibits the projected bandgap at the
$\bar{\Gamma}$-point (cf.\ Fig.~1(b)).
 As the (metal) electrons will actually have very small
scattering probability at the border of this SBZ, this fact does not
change the arguments given above for the decay of electrons excited
into the IS, but it complicates the search for the IS in the
calculated band structure.
 For that purpose we have analyzed the spatial probability distribution of
states at different $\mathbf{k}$-points along symmetrical
directions, averaged over the spatial coordinates $xy$ within the
supercell (Fig.~2).

Characteristic for the IS is a probability density along the surface
normal $\rho_z$ that resembles that of the
Shockley state of bare Ag(111) in the silver film.
 It has a pronounced maximum in the region of the metal-organic
interface and additional peaks around the center of the NTCDA
molecules (Fig.~2(c)). The dispersion of the IS reflects its
metallic character and can be described by a free-electron like
parabola (Fig.~2(a)). Within our accuracy, the effective mass is
independent of the geometry ($0.44\,m_e$ for flat and $0.45\,m_e$
for bent molecules) and close to that of the Shockley
state.\cite{Forster04jpcb}

 In contrast to the effective mass, the IS energy position at the
$\bar\Gamma$-point is very sensitive to the exact geometry of the
NTCDA adlayer (arrows in Fig.~2(a)).
 For the distance of 2.997~\AA, the bending only leads to a slight
upshift of 41~meV, from 243~meV to 284~meV. The reduction of the
distance by merely 5\% to 2.85~\AA, however, induces an upshift of
112~meV to 355 meV.
 The, as compared to NTCDA, higher IS energy of the commensurate PTCDA
layer can therefore to a large extent be explained by its slightly
smaller adsorption distance.

 Another important factor is the relative density of molecules per unit cell
area in general and carbon rings in particular.
 Keeping only one NTCDA molecule in the unit cell, brings the IS
energy $\sim100$~meV closer to the Fermi level.
 This finding explains why, in spite of the shorter adsorption
distance, the IS is observed at lower energy in the disordered, and
consequently less densely packed LT-PTCDA phase.
 Furthermore, the previously calculated IS energy for PTCDA, with
an adsorption distance of 2.85~\AA, is 469 meV at the
$\bar\Gamma$-point.\cite{Zaitsev10}
 Although this value, corresponding to the bent geometry, should
not be quantitatively compared with the 355~eV, obtained for flat
NTCDA at 2.85~\AA, the bending of the carboxyl groups is unlikely to
be the dominating factor for the larger upshift of the PTCDA-IS.
 Rather this comparison suggests that, together with the reduction
of the distance from the metal surface, the higher density of carbon
rings per unit cell area can account for the experimentally observed
energy difference of 200~meV between the commensurate NTCDA and
PTCDA layers.
 Both factors steepen the potential at the metal/organic
as compared to the metal/vacuum interface and thus cause the
Shockley state of the metal to shift to higher energies.

Whereas the bending of the carboxyl groups does not have a large
effect on the IS energy position, it significantly changes the
lateral probability distribution $\rho_{xy}$ of the wave function.
 In particular, the amplitude at the bridging oxygen atoms and at the
carbon atoms of the carboxyl groups is greatly enhanced as compared
to the flat molecule (Fig.~2(b)). In addition, $\rho_{xy}$ is
redistributed within the naphthalene core.
 Averaged over the whole SBZ, $\rho_z$ becomes larger on the
molecule and diminished in the metal (Fig.~2(c)). This explains the
longer lifetime of the IS of LT-PTCDA compared to the ordered phase
since NIXSW experiments have shown a significantly stronger bending
of the carboxyl groups of these molecules towards the
metal.\cite{Kilian08prl}


In conclusion, we investigated the formation mechanism of unoccupied
Shockley-derived metal-organic interface states. With adsorption of
the PTCDA and NTCDA molecules on Ag(111) the Shockley state of the
metal is shifted to energies above the Fermi level. The new energy
position is primarily influenced by the presence of a $\pi$-electron
system. It depends on adsorption distance and areal concentration of
carbon rings. In contrast, the spatial probability density of the state,
which has a pronounced maximum between the terminating silver atoms
and the carbon plane, is mainly determined by the molecular geometry.
Bending of the carboxylic end groups reduces the metallic
and enhances the molecular overlap.

Funding by the Deutsche Forschungsgemeinschaft and the Ikerbasque
Foundation is gratefully acknowledged.



\begin{thebibliography}{10}
\bibitem{fn1} Unlike in the IR+UV excitation scheme employed in
    Ref.\cite{Schwalb08prl},
   the image-potential states do not contribute to the measured
   2PPE intensity with Blue+UV excitation.
\bibitem{fn2} The peak shows an strong asymmetry towards higher
    energies, most likely due to contributions from stochastically formed
    small Herringbone areas. The evaluated position of the peak maximum
    shifts only little with emission angle,
    probably because parallel momentum in
    2PPE is not conserved for the disordered layer.
\end{thebibliography}

\end{document}